\documentclass[aps,twocolumn,showpacs,prl,amsmath,amssymb,floatfix,superscriptaddress]{revtex4-1}
\usepackage{amsmath,amssymb,graphicx,color}
\usepackage{bm}
\usepackage{bbm}
\usepackage[caption=false]{subfig}
\usepackage[usenames,dvipsnames]{xcolor}
\usepackage[colorlinks=true,citecolor=Cerulean,linkcolor=RubineRed,urlcolor=Cerulean]{hyperref}
\usepackage[normalem]{ulem}
\usepackage{soul,xcolor}
\newcommand{\mi}{\mathrm{i}} 

\usepackage{mathptmx} 

\begin{document}
\setstcolor{blue}
\title{Stabilization of Nonlinear Lattices: A Route to Superfluidity and Hysteresis}

\author{Gentaro Watanabe}\email{gentaro@zju.edu.cn}
\affiliation{Department of Physics and Zhejiang Institute of Modern Physics, Zhejiang University, Hangzhou, Zhejiang 310027, China}
\author{Yongping Zhang}\email{yongping11@t.shu.edu.cn}
\affiliation{Department of Physics, Shanghai University, Shanghai 200444, China}
\begin{abstract}
  The Bloch states of a Bose-Einstein condensates (BECs) in pure nonlinear lattices (NLs) are dynamically unstable, so that they cannot show superfluidity. We overcome this problem by finding that the two-component BECs in NLs can be stabilized by the coherent linear coupling. Furthermore, in the limit of the strong coherent coupling, the lowest Bloch band in the whole Brillouin zone can be dynamically stable. We also find a hysteretic behavior with loop-like structure in the Bloch band resembling the so-called ``swallowtail'' loop. The dynamical stabilization and hysteretic behavior can be observed in experiments by current technology.
\end{abstract}

\maketitle

\textit{Introduction.---} Understanding and controlling nonlinear phenomena is a profound issue in various areas of natural sciences. However, such programs often end up with difficulty because of the lack of controllability required to tune the system parameters. Recent development of the experimental techniques of cold atomic gases \cite{pethick_smith_book} and nano-photonics \cite{nano_photonics} is changing the situation, which allows us to realize highly controllable nonlinear systems. In cold atomic gases, nonlinearity emerges due to the presence of the superfluid order parameter and the strength of the nonlinearity can be controlled by a celebrated technique called the Feshbach resonance \cite{chin10}. Currently, even fine control of the strength with high spatial resolution at the submicron scale is possible \cite{yamazaki10} using the optical Feshbach resonances \cite{fedichev96}. In addition, various techniques allow optical control of the strength with low loss rate \cite{ciurylo05,zelevinsky06,enomoto08,bauer09a,bauer09b,yan13,fu13} leading to the longer lifetime of the sample up to of order $100$~ms \cite{blatt11,clark15,kim16}. In photonics, nonlinearity can be introduced by the Kerr media and recent nano-fabrication technique enables us to make nonlinear media with various fine structures such as nano-waveguides and photonic crystals \cite{photonic_crystals}.

Nonlinear lattices (NLs) are the novel setup recently realized in the above highly controllable nonlinear systems \cite{kartashov11}. There, the periodicity of the system is set by the nonlinear term, instead of the linear external potential unlike the ordinary crystalline lattices. In Bose-Einstein condensates (BECs) of ultracold atomic gases, this can be achieved by periodically modulating in space the coefficient of the nonlinear term, the $s$-wave scattering length, using, e.g., an optical Feshbach resonance \cite{fedichev96,yamazaki10}. For electromagnetic waves, NLs can be realized using photonic crystals with alternating layers of different nonlinear media \cite{nonlinlat_photonics}. There is on-going active research on various nonlinear objects such as solitons and vortices in NLs of these systems (see, e.g., \cite{kartashov11,nano_photonics} and references therein).

Nevertheless, the study of NLs is facing a crucial challenge: even at zero superflow, we cannot keep stationary periodic states in NLs because of the dynamical instability unless we introduce a sufficiently large uniform repulsive interaction coefficient \cite{zhang13}. Therefore, almost all experimental studies of the NL so far are limited to the above-mentioned localized nonlinear objects. Stabilizing extended periodic states in pure NLs without uniform interaction coefficient to achieve superfluidity in this system stands out as an open problem.

In this work, we overcome this challenge using two-component BECs. We find that the NLs can be stabilized by an interspecies, coherent linear coupling. In particular, the lowest branch can be dynamically stable in the whole Brillouin zone (BZ) in the limit of strong coherent coupling. We also find that this system can show a hysteretic behavior with a loop-like structure in the Bloch band which resembles the so-called ``swallowtail'' loop \cite{wu02,diakonov02,mueller02,machholm03,seaman05,chen11,stfermi,hui12,eckel14,yu15,koller16}.

\textit{Setup.---} We consider two-component BECs (labeled by component $a$ and $b$) in NLs generated by the spatially periodic intraspecies interaction strength $g_{a}$ and $g_{b}$. For simplicity, we assume that the NLs are one-directional, which we take to be in the $x$ direction. We further assume that the period $d$ of the NL and the mass $m$ of the atoms, which is set to unity, are the same for component $a$ and $b$. We introduce the coherent linear coupling between the two NLs whose energy functional $E_{\rm CLC}$ is given by
\begin{equation}
  E_{\rm CLC} = \frac{\Omega_0}{2} \int d{\bf r}\, (\Psi_a^*\, \Psi_b + \mathrm{c.c.})\, ,
\label{eq:hclc}
\end{equation}
where $\Omega_0$ is the Rabi frequency, which is set to be real, and $\Psi_\sigma$'s are the condensate wave function of component $\sigma=\{a, b\}$.

This system can be described by the time-dependent Gross-Pitaevskii (GP) equation:
\begin{align}
  \mi \partial_t
  \begin{pmatrix}
    \Psi_a\\ \Psi_b
  \end{pmatrix}
  = \left[
  \begin{pmatrix}
    H_a & 0 \\
    0 & H_b
  \end{pmatrix}
  + \frac{\Omega_0}{2} \sigma_x
  \right]
  \begin{pmatrix}
    \Psi_a\\ \Psi_b
  \end{pmatrix}
  \equiv H_{\rm GP}\psi\label{eq:gpemat}
\end{align}
with $\psi \equiv (\Psi_a, \Psi_b)^\intercal$, $\sigma_x$ being the Pauli matrix, and
\begin{align}
  H_\sigma \equiv -\frac{1}{2} \partial_x^2 + g_{\sigma}(x) |\Psi_{\sigma}|^2 + g_{ab} |\Psi_{\bar{\sigma}}|^2\,,
\end{align}
where $\bar{\sigma} = \{b, a\}$ for $\sigma = \{a, b\}$. Here, $g_{\sigma}(x) = g_{\sigma}^{(0)} + \frac{\Delta g_{\sigma}}{2} \cos(2k_0x)$ with $k_0=\pi/d$, and $g_{ab}$ is the interspecies interaction strength. The two NLs are in-phase when $\Delta g_{a}$ and $\Delta g_{b}$ have the same sign, and are out-of-phase when they have the opposite sign. Since one of our main purposes is to show that the pure NLs without the uniform component of the interaction strength can be stabilized, we set $g_{\sigma}^{(0)} = 0$ hereafter.

\textit{Discrete model.---} This setup could be mapped to a simplified discrete model \cite{nonlinlatrev,pd_nonlinlat}: reducing the system with a spatially periodic interaction strength in the continuum representation to a discrete representation by sampling just two points per period of the interaction strength (the maxima and minima of the interaction strength). Thus in this discrete model, the spacing between two neighboring sites is $\tilde{d} = d/2$. In this representation, the on-site interaction parameter alternates between $U_\sigma$ and $-U_\sigma$ ($\sigma=\{a,b\}$) at the adjacent sites. Assuming that the hopping parameter $K$ is equal between component $a$ and $b$, the Hamiltonian is written as
\begin{align}
  H =& \sum_{\sigma=\{a,b\}} \sum_j \left[\, -K (c_{\sigma, j}^*\, c_{\sigma, j+1} + \mathrm{c.c.}) + (-1)^j\,\, \frac{U_\sigma}{2} |c_{\sigma, j}|^4\, \right] \nonumber\\
  & + \frac{U_{ab}}{2} \sum_j |c_{a, j}|^2\, |c_{b, j}|^2 + \frac{\Omega}{2} \sum_j (c_{a, j}^*\, c_{b, j} + \mathrm{c.c.}),
\label{eq:hamil}
\end{align}
where $c_{\sigma, j}$ is the amplitude of component $\sigma$ at site $j$, $U_\sigma$ is the on-site interaction parameter for species $\sigma$, $U_{ab}$ is the inter-component interaction parameter, and $\Omega$ is the Rabi coupling constant. The discrete model has been used in the previous studies of superflow in NLs \cite{nonlinlatrev,pd_nonlinlat}. These studies have demonstrated that the discrete model can capture well the qualitative properties of the extended states of BECs in NLs including their stability properties. Furthermore, we have confirmed that the discrete model can describe well the key features of the present system obtained from the full continuum model as well.

Equations of motion (EOMs) in the discrete model are given by the Euler-Lagrange equations for $c_{\sigma, j}$ with the Lagrangian $\mathcal{L} = \sum_{\sigma} \sum_j \left[ \frac{\mi}{2} (c_{\sigma, j}^*\, \partial_t c_{\sigma, j} - c_{\sigma, j}\, \partial_t c_{\sigma, j}^*) \right] - H$\, with $H$ given by Eq.~(\ref{eq:hamil}). They read
\begin{align}
  \mi \dot{c}_{\sigma, j} =& -K (c_{\sigma, j+1} + c_{\sigma, j-1}) + (-1)^j\,\, U_{\sigma} |c_{\sigma, j}|^2\, c_{\sigma, j} \nonumber\\
  & + \frac{U_{ab}}{2} |c_{\bar{\sigma}, j}|^2\, c_{\sigma, j} + \frac{\Omega}{2} c_{\bar{\sigma}, j}\, .\label{eq:eom_discrete}
\end{align}

In this work, we consider periodic states in which $c_{\sigma, j}$'s are in the Bloch form: $c_{\sigma, j} = h_{\sigma, j}\, \exp{(\mi k j \tilde{d}\,)}$, where $k$ is the quasi-wavenumber of the bulk superflow \cite{note:equal_k}, which is defined within the first BZ $|k| \le k_0$, and $h_{\sigma, j}$'s are the complex amplitudes with period $2\tilde{d}$, i.e., $h_{\sigma, j} = h_{\sigma, j+2}$. Since we have real and imaginary parts for four amplitudes $h_{\sigma, 1}$ and $h_{\sigma, 2}$ $(\sigma=\{a, b\})$, we have 8 variables in total. However, we have a constraint on the total number of particles $\nu$ per unit cell, which consists of two sites, and we can set the phase of one of the amplitudes (e.g., the phase of $h_{a, 1}$) to be zero without loss of generality, thus the number of independent variables is reduced to~$6$.
Stationary solutions of $c_{\sigma, j}=c_{\sigma, j}^{(0)}$ are obtained by extremizing the Hamiltonian (\ref{eq:hamil}) under the normalization condition, $\sum_{\sigma}(|h_{\sigma,1}|^2 + |h_{\sigma,2}|^2) = \nu$, with respect to the six independent variables.

The stability of superfluidity can be judged by whether perturbations added to the stationary state grows in time, using the linear stability analysis (e.g., Ref.~\cite{wu03}, Sec.~14.3 in Ref.~\cite{pethick_smith_book}, and Sec.~5.6 in Ref.~\cite{pitaevskii_stringari_book}). We first linearize the EOMs (\ref{eq:eom_discrete}) in terms of the perturbations $\delta c_{\sigma, j}(t)$ around the stationary solution $c_{\sigma, j}^{(0)}$. We consider the following general form of the perturbations for the periodic system, $\delta c_{\sigma, j} = e^{\mi kj\tilde{d} - \mi \mu t} ( u_{\sigma, j}\, e^{\mi qj\tilde{d} - \mi \omega t} + v_{\sigma, j}^*\, e^{-\mi qj\tilde{d} + \mi \omega t} )$, where $q$ is the quasi-wavenumber of quasiparticles, $\mu$ is the chemical potential of the stationary state, and the Bogoliubov quasiparticle amplitudes $u_{\sigma, j}$ and $v_{\sigma, j}$ have the same periodicity (period $2\tilde{d}$) as the stationary solutions. Substituting $c_{\sigma, j}(t)=c_{\sigma, j}^{(0)} + \delta c_{\sigma, j}(t)$ into the linearized EOMs, we obtain an eigenvalue equation $\mathcal{M}\mathbf{u} = \omega\mathbf{u}$ with $\mathbf{u}^\intercal = (u_{a, 1}, v_{a, 1}, u_{a, 2}, v_{a, 2}, u_{b, 1}, v_{b, 1}, u_{b, 2}, v_{b, 2})$ and $\mathcal{M}$ being an $8\times 8$ matrix.

It is noted that the matrix $\mathcal{M}$ is not Hermitian and thus its eigenvalues $\omega$ are complex in general. The condition for dynamical stability is that all the eigenvalues are real for any $q$; otherwise, the perturbation corresponding to an eigenvalue with nonzero imaginary part grows exponentially in time. A mode corresponding to $\omega$ with the maximum nonzero absolute value of the imaginary part is the fastest growing mode.

In the following, we focus on the case with $|U_a|=|U_b|$.

\begin{figure}[t!]
\centering
\includegraphics[height=4.1cm]{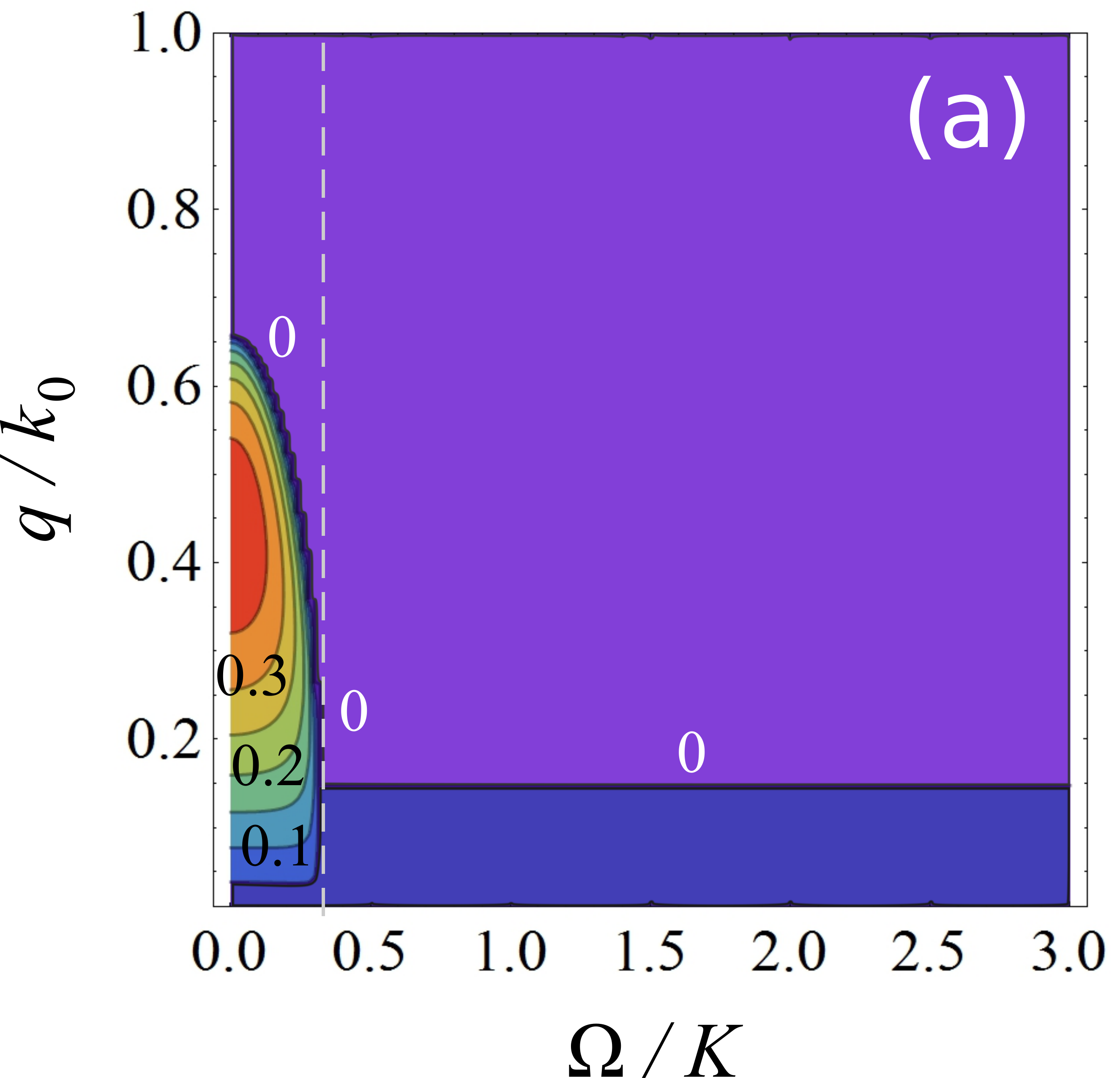}
\includegraphics[height=4.1cm]{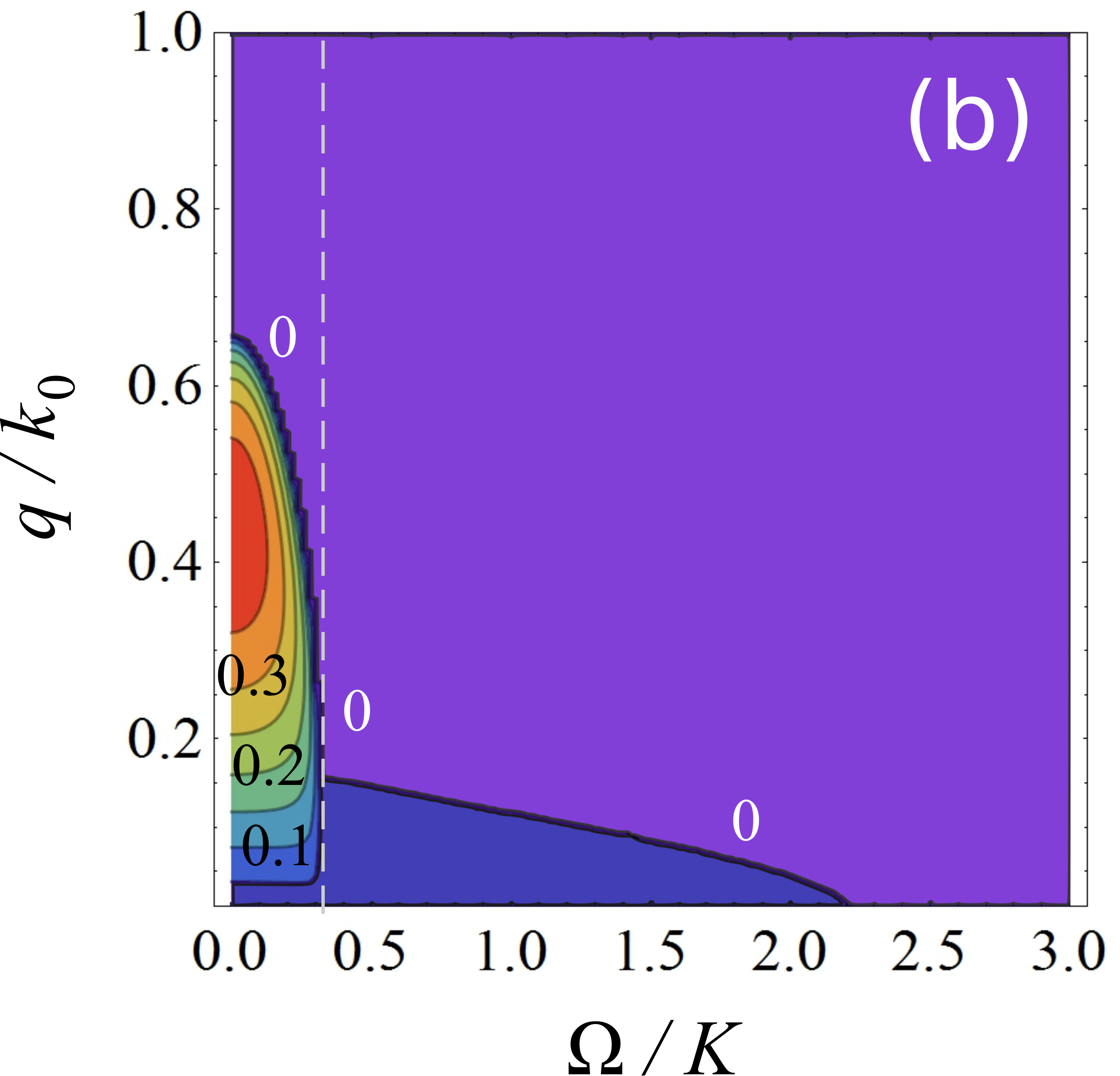}
\includegraphics[height=4.1cm]{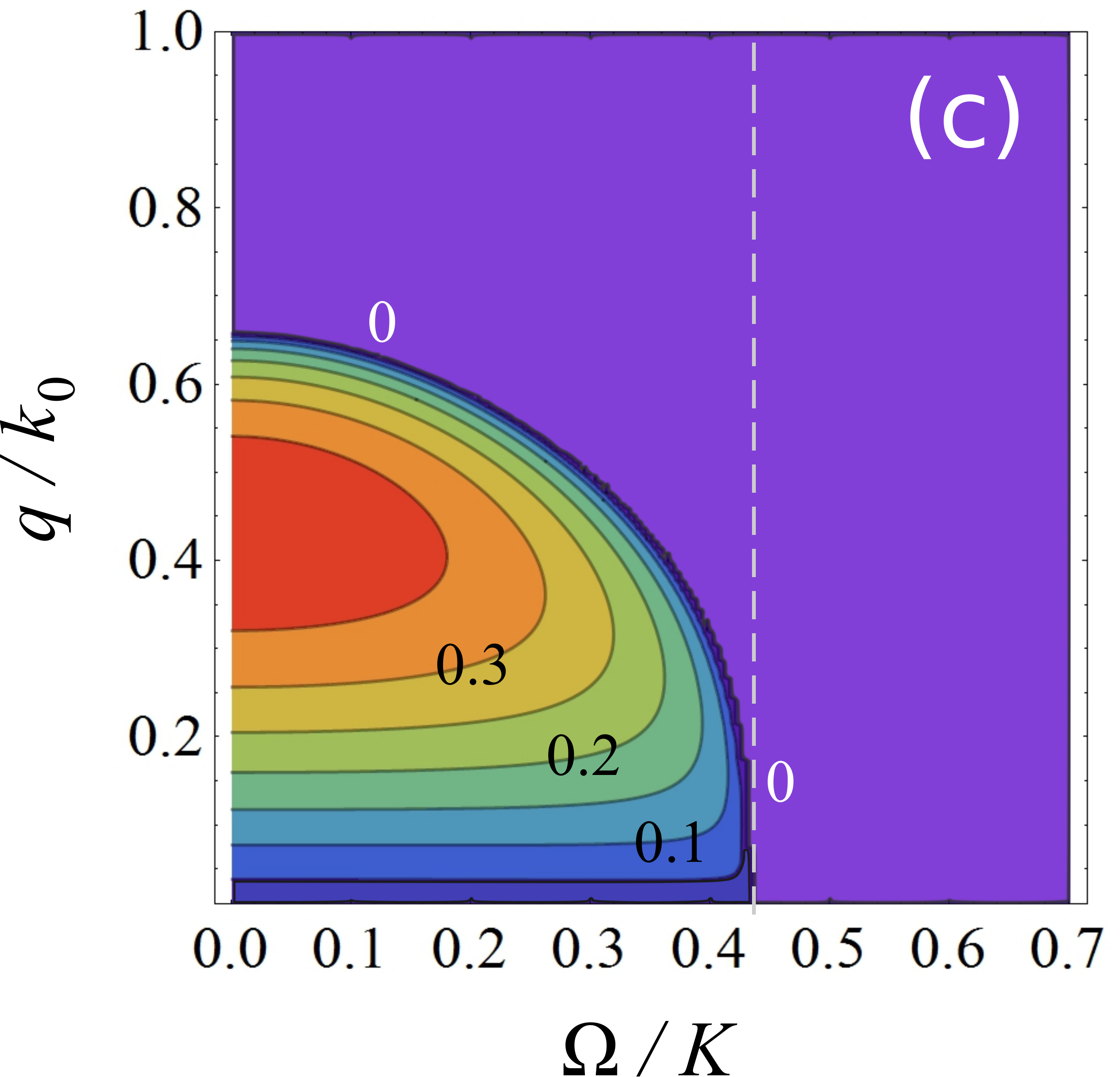}
\includegraphics[height=4.1cm]{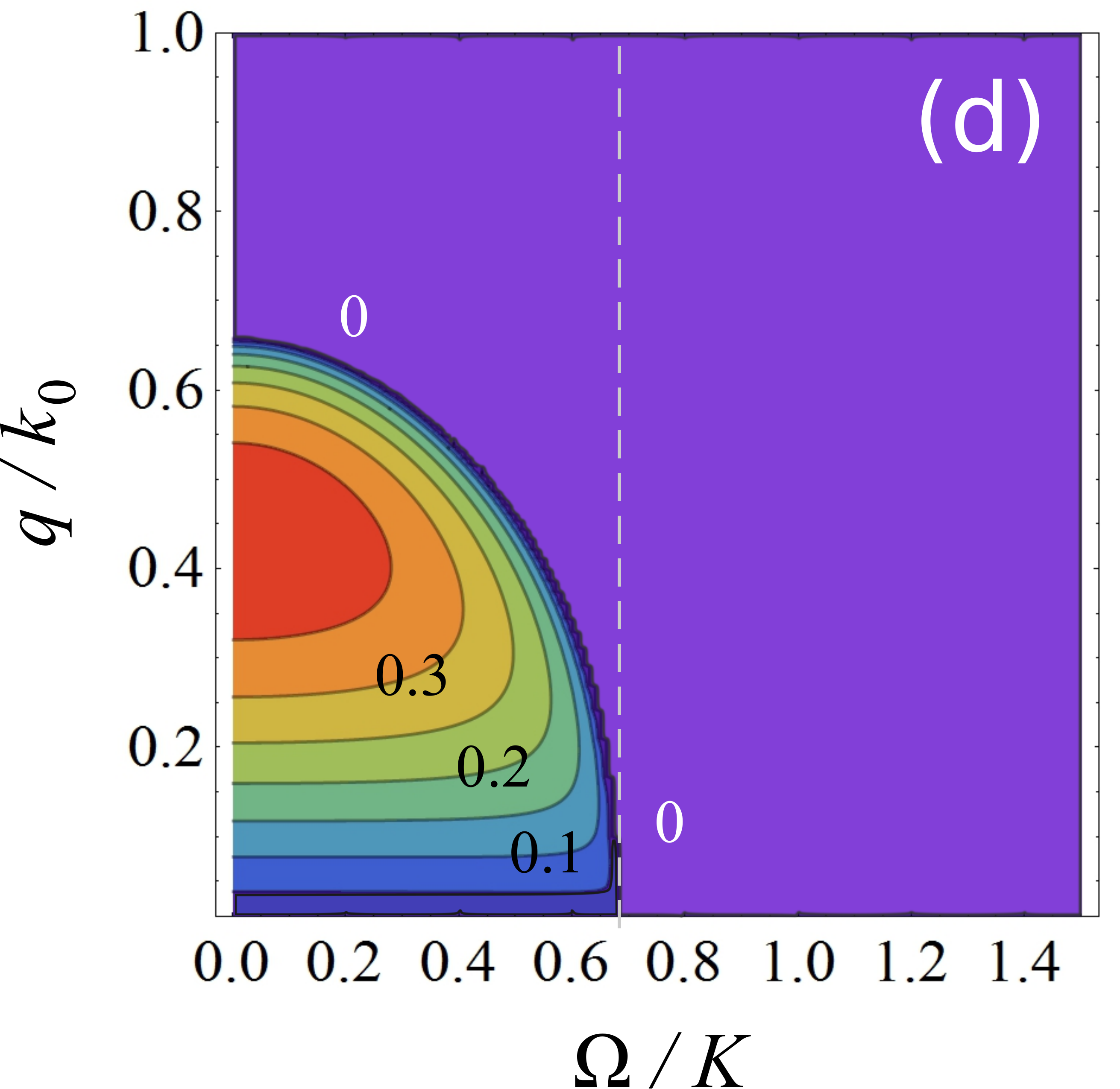}
\caption{Dynamical stability diagrams of the lowest stationary states at $k=0$ in the coherently coupled NLs. The top two panels are for the in-phase (a) and out-of-phase (b) cases at $U_{ab}\nu/K =0.5$. The bottom panels are for the out-of-phase case at $U_{ab}\nu/K =1$ (c) and $2$ (d); for these parameter values, results of the in-phase case are almost identical to that of the out-of-phase case. Here, we set $|U_{\sigma}|\nu/K =1$. The contours show the growth rate of the fastest growing mode, i.e., the maximum absolute value of the imaginary part of the eigenvalues of the matrix $\mathcal{M}$ in units of $K$. The vertical dashed line shows the boundary of $\Omega$ below (above) which the lowest state is a polarized (unpolarized) one.
}
\label{fig:stability_diagram_discrete}
\end{figure}

\textit{Stabilization of static systems.---} First, we discuss the dynamical stability of the static case without superflow (i.e., $k=0$). The stationary periodic states are obtained by numerically solving the time-independent GP equation for the Hamiltonian (\ref{eq:hamil}). There are two possible types of solutions both in the in-phase and out-of-phase cases: 1) polarized states with $\nu_a \ne \nu_b$ and 2) unpolarized states with $\nu_a=\nu_b$, where $\nu_a$ and $\nu_b$ are the number of particles of $a$- and $b$-component per unit cell. The polarized and unpolarized states reduce to the immiscible (phase separated) and miscible states, respectively, in a mixture of two-component BECs corresponding to the limit of $\Omega=0$. With increasing $\Omega$, the two components are coherently ``mixed'' and thus the unpolarized state starts to be favored over the polarized state at sufficiently large $\Omega$.

Figure~\ref{fig:stability_diagram_discrete} shows the results of the linear stability analysis for the lowest periodic stationary state at $k=0$ plotting the maximum absolute value of the imaginary part of the eigenvalues of the matrix $\mathcal{M}$. In the violet regions, all the eigenvalues $\omega$ are real; if all of $\omega$ are real at any quasimomentum $q$ of excitations for a given $\Omega$, the system is dynamically stable at this $\Omega$. In both the cases of in-phase and out-of-phase NLs, when $\Omega$ is sufficiently small, the lowest state at $k=0$ is a polarized state, while above some threshold value of $\Omega$ (shown by the vertical white dashed lines in Fig.~\ref{fig:stability_diagram_discrete}), the polarized state no longer exists at $k=0$ and an unpolarized state becomes the lowest state. This critical value of $\Omega$ increases with increasing $U_{ab}\nu/K$ for a fixed $U_{\sigma}\nu/K$.

Figures~\ref{fig:stability_diagram_discrete}(b)--\ref{fig:stability_diagram_discrete}(d) demonstrate that the NLs could be stabilized by the linear coherent coupling with large $\Omega$. A striking difference between the in-phase and the out-of-phase cases is that the out-of-phase NLs can always be stabilized by setting $\Omega$ large enough while the in-phase NLs cannot be for smaller values of $U_{ab}$ \cite{note:inst_inphase}. This difference can be clearly seen from the top two panels [(a): in-phase, (b): out-of-phase]. For larger values of $U_{ab}$, an unpolarized state becomes dynamically stable when it becomes the lowest state due to the disappearance of a polarized state at $k=0$ [Figs.~\ref{fig:stability_diagram_discrete}(c) and \ref{fig:stability_diagram_discrete}(d)]. In this regime, the stability diagrams of the in-phase and the out-of-phase cases are almost identical.

\begin{figure}[tb!]
\centering
\includegraphics[width=0.495 \columnwidth]{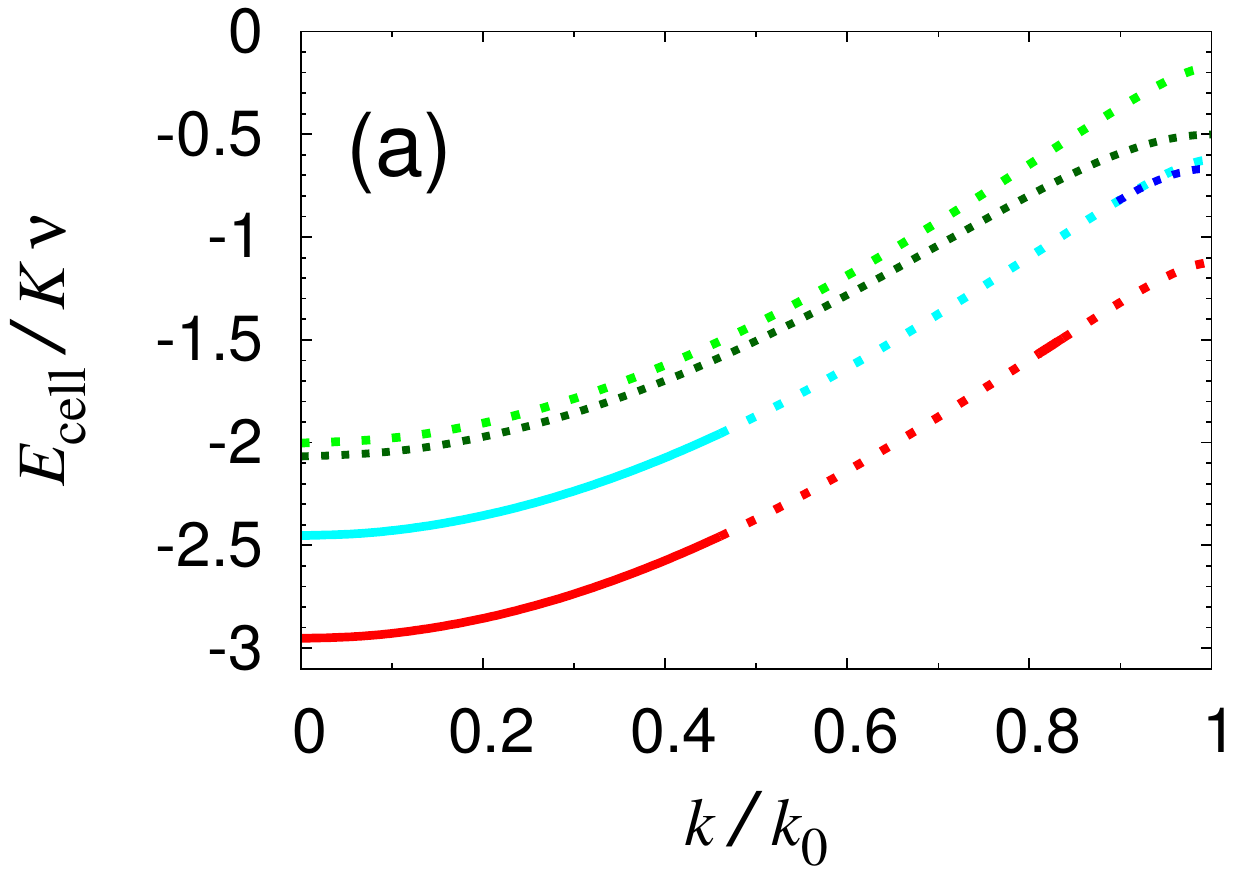}
\includegraphics[width=0.495 \columnwidth]{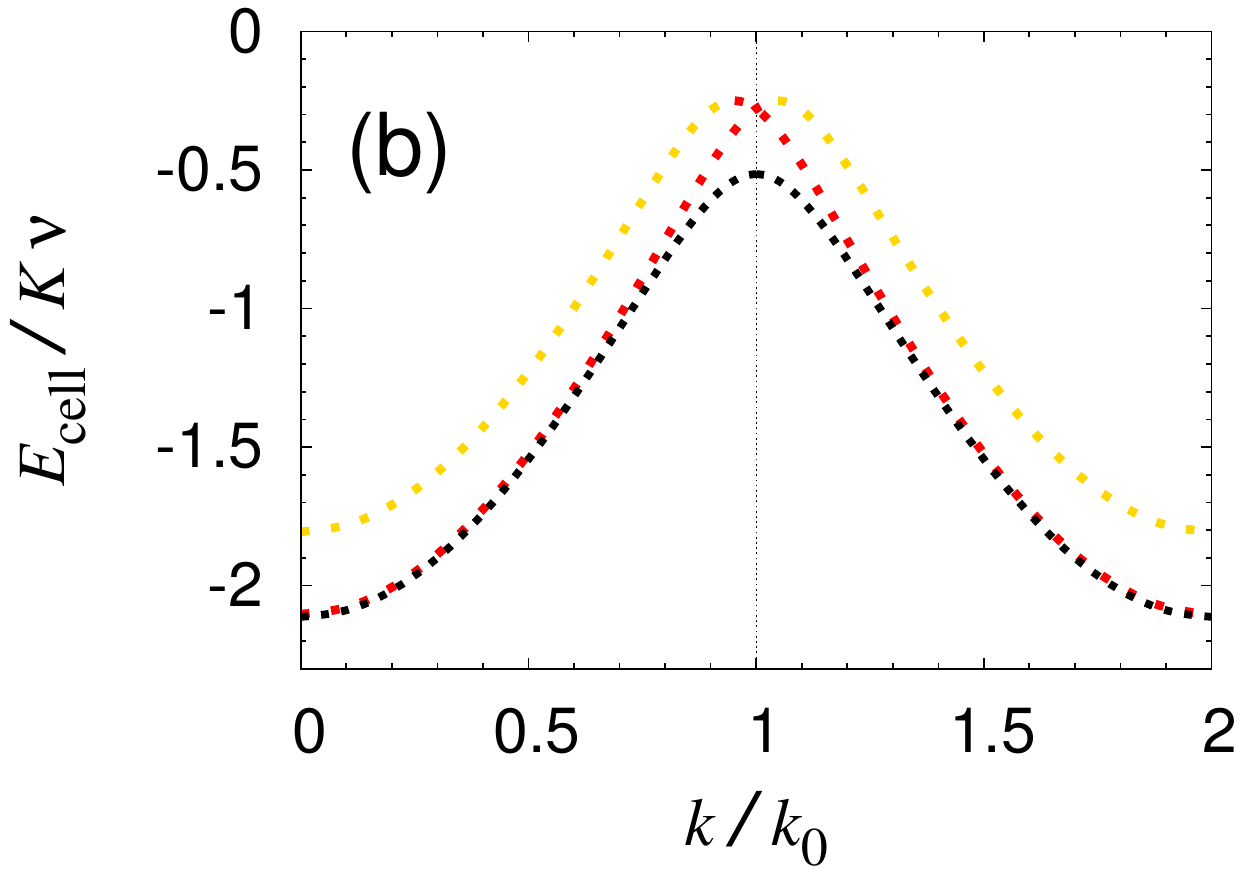}
\includegraphics[width=0.495 \columnwidth]{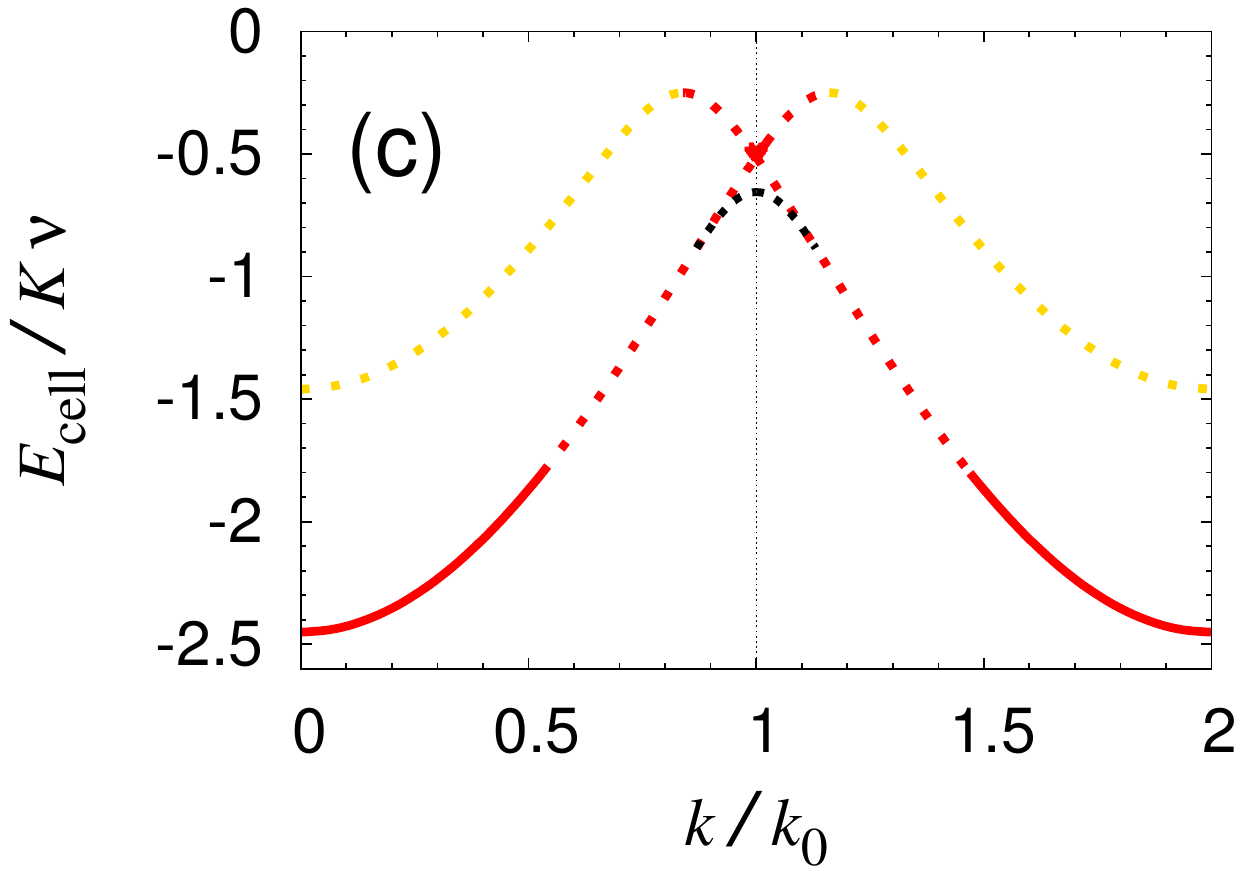}
\includegraphics[width=0.495 \columnwidth]{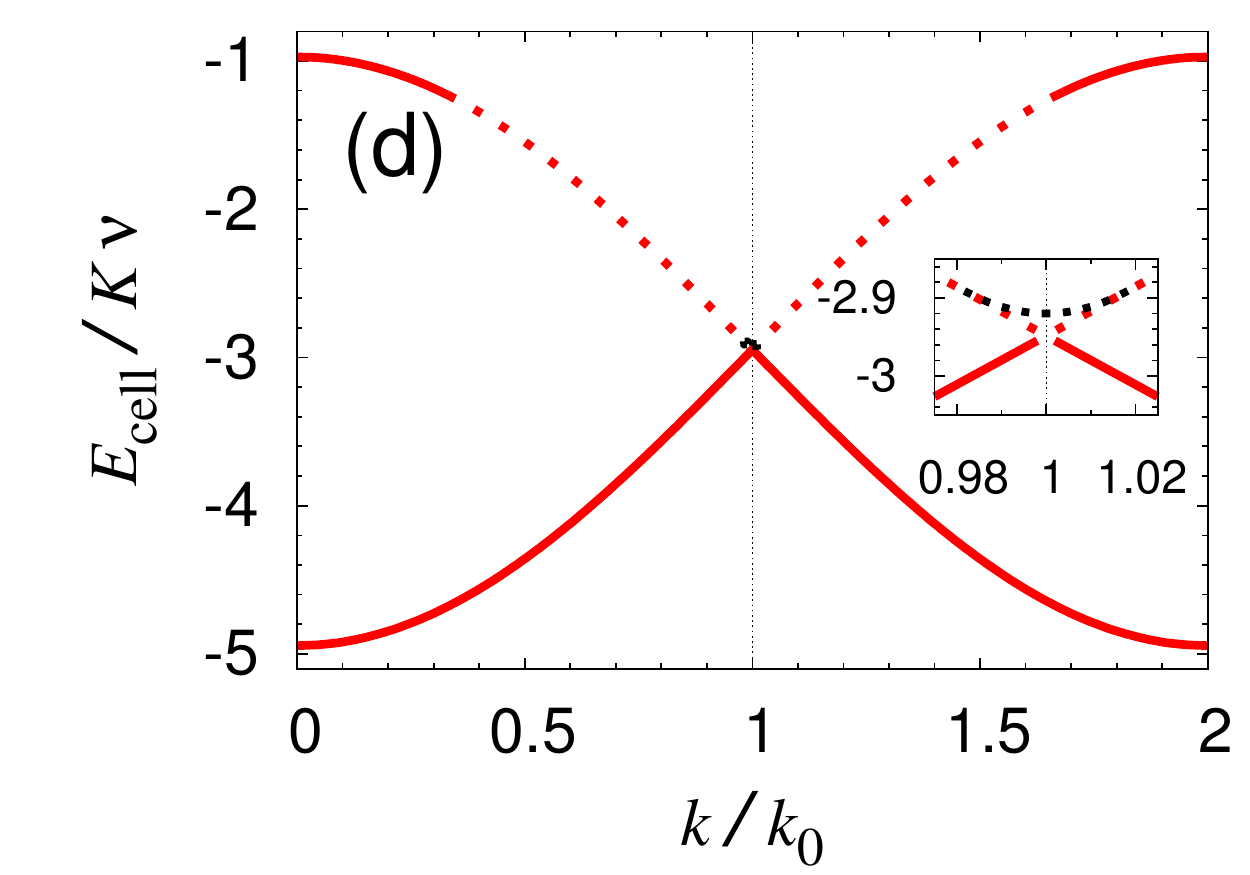}
\caption{Energy band for the in-phase (a) and out-of-phase (b)--(d) NLs: Energy per particle $E_{\rm cell}/\nu$ in units of $K$ as a function of $k$ for various values of $\Omega/K$. Here we set $|U_\sigma|\nu/K=U_{ab}\nu/K=1$. In (a) for the in-phase case, the green and the dark green lines are for $\Omega/K=0.1$, the cyan and the blue lines are for $\Omega/K=1$, and the red line is for $\Omega/K=2$. The dark green and the blue (densely dotted) lines are for the polarized branch and the others are for the unpolarized branch. Among the three panels for the out-of-phase case, $\Omega/K=0.3$ (b), $1$ (c), and $6$ (d). In these three panels, the black (densely dotted) line is for the polarized branch, the red line is for the lowest unpolarized branch, and the yellow line is for the second lowest unpolarized branch. The inset of (d) is a magnification around $k/k_0=1$. In all four panels, the solid line corresponds to the dynamically stable region of $k$ and the dotted line (both the densely and loosely dotted) to the dynamically unstable one. The first BZ is $|k|\le k_0$ and the second BZ is $k_0 < |k| \le 2k_0$.
}
\label{fig:ek}
\end{figure}

\textit{Energy band structure.---} We next discuss the energy band structure of the coherently coupled NLs. Figure \ref{fig:ek} shows the energy per particle $E_{\rm cell}/\nu$ for the in-phase (a) and the out-of-phase case (b)--(d) calculated from Eq.~(\ref{eq:hamil}) as a function of the quasi-wavenumber $k$ of the superflow. For small $\Omega/K$, in both the cases of in-phase and out-of-phase NLs, the polarized branch extends the whole first BZ [the dark green line in (a) and the black line in (b)] and is lower in energy compared to the unpolarized branch [the light green line in (a) and the red line in (b)], so that the polarized branch is the lowest. For larger $\Omega/K$, the polarized states start to disappear from the BZ center ($k=0$) and the region of $k$ of the polarized branch becomes narrower [the blue line in (a) and the black line in (c)]. Consequently, the unpolarized state becomes the lowest in energy in the region of $k$ where the polarized states have disappeared. Note that the unpolarized states can be dynamically stable at sufficiently large $\Omega/K$ while the polarized states are always dynamically unstable. In the in-phase case, the polarized branch completely disappears at sufficiently large $\Omega/K$ while, in the out-of-phase case, it remains to exist around the BZ edge at $k/k_0=1$ and finally becomes higher in energy compared to the lowest unpolarized branch [the inset of (d)]. Remarkably, in the limit of large $\Omega/K$, the lowest branch in the out-of-phase NLs is dynamically stable in the whole BZ as can be seen from the red solid line in (d).

In out-of-phase NLs, the lowest unpolarized branch shows the hysteretic behavior such that the dispersion with the positive slope persists beyond the first BZ edge at $k=k_0$ [the red lines in (b)--(d)] unlike the standard form of the Bloch band as in the in-phase case [panel (a)] which is periodic with period $2k_0$ and has zero slope at the first BZ edge. Together with the polarized branch [the black lines in (b)--(d)], which is periodic with period $2k_0$, the hysteretic unpolarized branch forms a loop-like band structure similar to the so-called ``swallowtail'' loop \cite{wu02,diakonov02,mueller02,machholm03,seaman05,chen11,stfermi,hui12,eckel14,yu15,koller16}. What is different from the swallowtail loop in the ordinary periodic potential is that the hysteresis appears at infinitesimally small values of the coefficients $|U_{\sigma}|\nu/K$ and $U_{ab}\nu/K$ of the nonlinear terms for any given nonzero $\Omega$.

Interestingly, the lowest and the second lowest unpolarized branches connect with each other as if they as a whole look like a single band with an enlarged BZ of $|k|\le 2k_0$ [the red and the yellow lines in Figs.~\ref{fig:ek}(b) and \ref{fig:ek}(c)]. This is because $|\Psi_a(x)| = |\Psi_b(x \pm d/2)|$ in the unpolarized states in the out-of-phase NLs with $|\Delta g_a|=|\Delta g_b|$ (i.e., $|U_a|=|U_b|$), and thus $H_{\rm GP}$ in Eq.~(\ref{eq:gpemat}) is invariant under the combined operation of the translation by half of the lattice constant and $\sigma_x$ operation. At sufficiently large $\Omega/K$, the lowest unpolarized branch extends over the whole first and second BZs and, as a consequence, the second lowest unpolarized branch disappears [panel (d)].

The physical mechanism of the emergence of the hysteresis in the out-of-phase NLs can be understood as follows. Introducing the phase $\theta_{\sigma}$ of the condensate wave function $\Psi_{\sigma}$ as $\Psi_{\sigma} = \sqrt{n_{\sigma}}\, e^{\mathrm{i} \theta_{\sigma}}$, $E_{\rm CLC}$ can be written as $E_{\rm CLC} = \Omega_0 \int d{\bf r}\, \sqrt{n_a n_b}\,\, \cos{(\theta_a - \theta_b)}$. Since $|\theta_a - \theta_b| = \pi$ in the lowest unpolarized branch, $E_{\rm CLC} = -\Omega_0 \int d{\bf r}\, \sqrt{n_a n_b}$\,. Therefore, $E_{\rm CLC}$ tends $n_a$ and $n_b$ to be equal at every site when $\Omega_0$ is large and positive: Namely, $E_{\rm CLC}$ with positive $\Omega_0$ has an effect of locking densities of $a$- and $b$-components with each other. In the out-of-phase NLs, this effect prevents the condensate wave functions from having a node, which is necessary to render the slope of the dispersion to vanish at the first BZ edge, and the hysteresis appears as a result.

\textit{Experimental feasibility.---} Finally, we discuss the experimental feasibility of our system using the realistic full continuum model given by Eq.~(\ref{eq:gpemat}). Here we take parameter values from the experiment of Ref.~\cite{yamazaki10}. Following this experiment, we consider $^{174}$Yb atoms in NLs created by the lattice laser with the wavelength $\lambda=555.8$ nm (i.e., the lattice constant $d=\lambda/2$ and $k_0= 2\pi/\lambda$). We set the variation of the intraspecies scattering length $\Delta a_\sigma$ of species $\sigma$ as $|\Delta a_\sigma| = |\Delta a_a| = |\Delta a_b| = 20$ nm, the interspecies scattering length $a_{ab}=5.55$ nm, and the average atom density $n=1.5 \times 10^{14}$ cm$^{-3}$. Therefore, the parameters in Eq.~(\ref{eq:gpemat}) are $|\Delta g_\sigma| n/E_0 \equiv (4\pi |\Delta a_\sigma|/m)n/E_0 = 0.3$ and $g_{ab}n/E_0 \equiv (4\pi a_{ab}/m)n /E_0 = 0.082$ with the energy unit $E_0\equiv k_0^2/m = 2\pi \times 7.4$ kHz.

\begin{figure}[tb!]
\centering
\includegraphics[width= 0.93 \columnwidth]{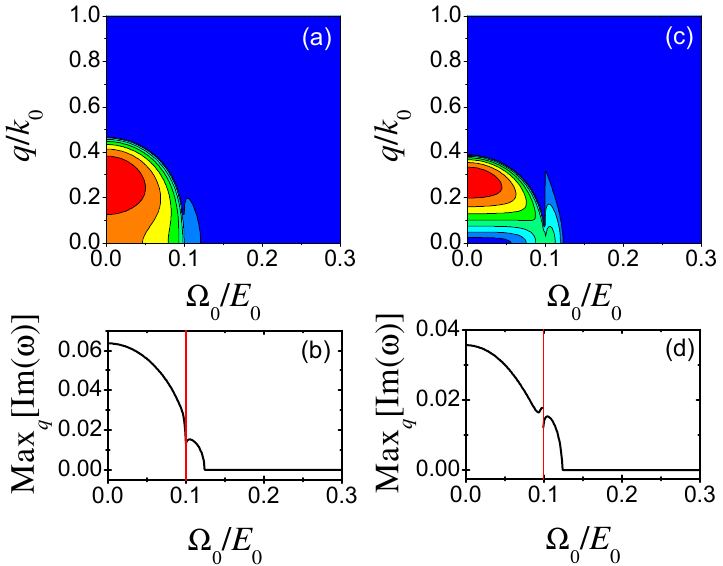}
\caption{Dynamical stability diagrams of the lowest stationary states at $k=0$ for the continuum model [(a) and (b): in-phase case; (c) and (d): out-of-phase case]. Similarly to Fig.~\ref{fig:stability_diagram_discrete}, (a) and (c) show the growth rate of the fastest growing mode in units of $E_0$, and (b) and (d) show its maximum value with respect to $q$. The red vertical line in (b) and (d) shows the boundary of $\Omega_0$ below (above) which the lowest state is a polarized (unpolarized) one.
}
\label{fig:cont}
\end{figure}

The dynamical stability diagrams for these parameter values are shown in Fig.~\ref{fig:cont}. We see that the NLs become dynamically stable for $\Omega_0/E_0 \agt 0.1$ in both the in-phase and out-of-phase cases. This magnitude of the Rabi frequency is easy to realize in current experiments.

\begin{figure}[t!]
\centering
\includegraphics[width= 0.66 \columnwidth]{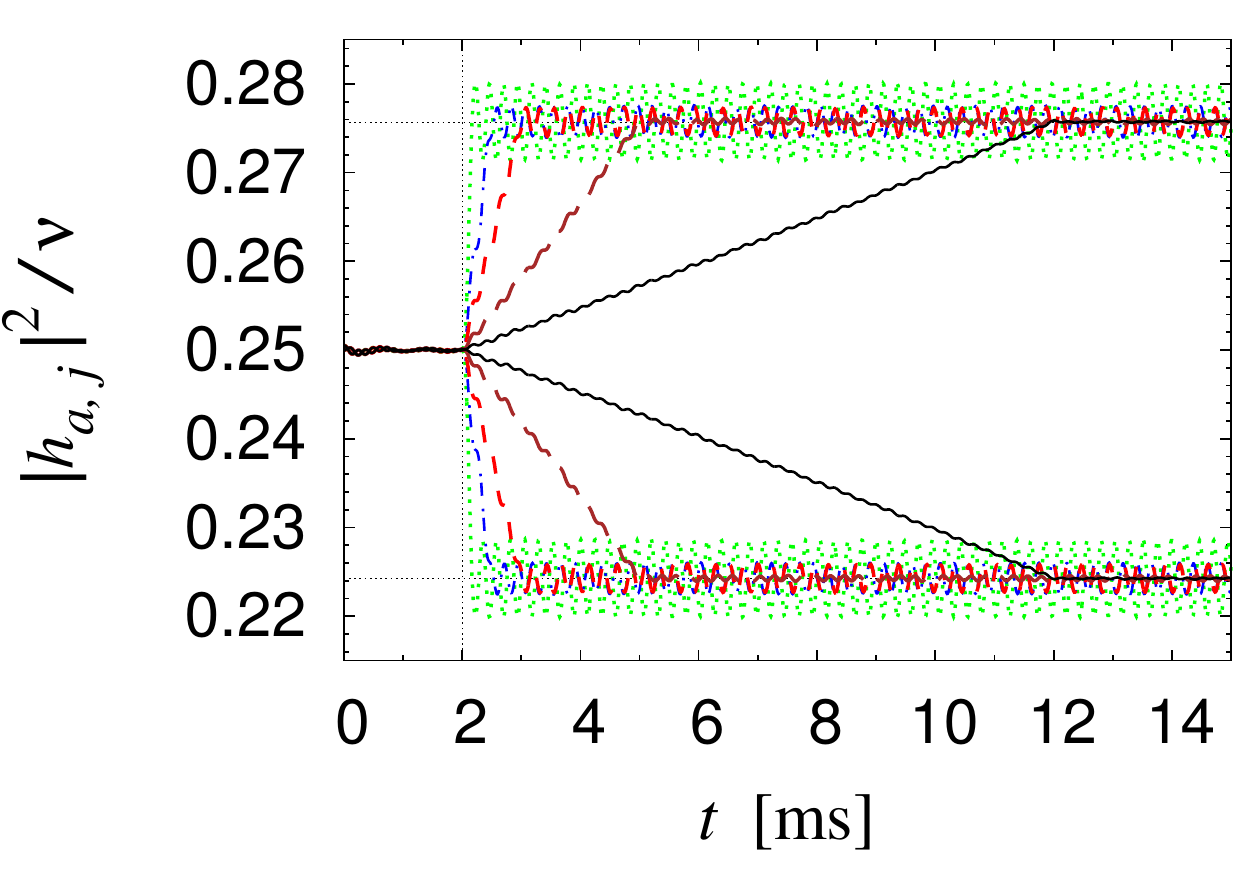}
\caption{Time evolution of the population of component $a$ at the first and the second sites when the out-of-phase NLs are linearly ramped. Here we use NLs with 100 unit cells with the periodic boundary conditions. The ramping time $T_{\rm ramp}$ is $0.2$ ms (green dotted), $0.5$ ms (blue dashed-dotted), $1$ ms (red short dashed), $3$ ms (brown long dashed), and $10$ ms (black solid).
}
\label{fig:ramp}
\end{figure}

In addition, stable NLs can be adiabatically prepared within a reasonable time scale of $\agt 1$ ms. Figure~\ref{fig:ramp} shows the time evolution of the population obtained from the discrete model when the NLs are linearly ramped with various ramping time $T_{\rm ramp}$. Parameters of the discrete model ($U_\sigma\nu/K$, $U_{ab}\nu/K$, $\Omega/K$, and $K/E_0$) corresponding to the above realistic setup can be uniquely determined by fitting the intraspecies and interspecies interaction energies, $E_{\rm CLC}$, and the effective mass of the continuum model. Results shown in Fig.~\ref{fig:ramp} corresponds to $\Omega_0/E_0=0.2$ \cite{note:parameters}. We consider two protocols to ramping the pure NLs with $g_\sigma^{(0)}=0$: 1) starting from a uniformly repulsive intraspecies interaction with the same value as $g_{ab}$ and 2) starting from no intraspecies interaction. While Fig.~\ref{fig:ramp} shows the result of the first protocol to realize the out-of-phase NLs, there is no significant difference between the results of the two protocols. According to Fig.~\ref{fig:ramp}, the case of $T_{\rm ramp} \simeq 1$ ms is already colse to the adiabatic limit. The results for ramping the in-phase NLs are also similar.


\textit{Summary and conclusion.---}
We have found that the dynamical instability of the BEC in NLs can be overcome using the coherent linear coupling between two components. Especially, in out-of-phase NLs, the lowest band in the whole BZ becomes dynamically stable in the limit of the strong coherent coupling. In addition, in out-of-phase NLs, the coherent linear coupling can generate the exotic energy band structure due to the hysteresis. Since the stabilization of NLs have been shown to be realized by current technology, further research on the superfluidity, transport properties, etc. and various applications of the NLs are awaited.

\begin{acknowledgments}
We thank B. Prasanna Venkatesh and Qijin Chen for discussions. G.W. was supported by NSF of China (Grant No. 11674283), by the Fundamental Research Funds for the Central Universities (2017QNA3005), by the Zhejiang University 100 Plan, and by the Thousand Young Talents Program of China. Y.Z. was supported  by the NSF of China (Grant No. 11774219), by the Thousand Young Talents Program of China, and by the Eastern Scholar and Shuguang (17SG39) Program of Shanghai.
\end{acknowledgments}


\begin{thebibliography}{99}
\bibitem{note:equal_k} Here we have assumed that the superflows of component $a$ and $b$ have the same $k=k_a=k_b$, which is required in order that the coherent coupling term of the Hamiltonian to be periodic. For $k_a\neq k_b$ except for $k_a-k_b = n\pi/\tilde{d}$ with integer $n$, we cannot have a periodic stationary state with period $d=2\tilde{d}$.
\bibitem{note:inst_inphase} This would be related to the dynamical instability of the unpolarized state in uniform systems triggered by a density mode at small values of intra and interspecies interaction strengths \cite{uniformsys}.
\bibitem{note:parameters} Parameters of the discrete model corresponding to $|\Delta g_\sigma| n/E_0 = 0.3$, $g_{ab}n/E_0 = 0.082$, and $\Omega_0/E_0=0.2$ are $|U_\sigma|\nu/K=1.07$ ($1.03$), $U_{ab}\nu/K=3.25$ ($3.21$), $\Omega/K=1.98$ ($1.97$), and $K/E_0=0.102$ ($0.102$) for the out-of-phase (in-phase) case.
\end{thebibliography}
\end{document}